\documentclass[12pt,preprint]{aastex}
\shorttitle{The Companion to NLTT 11748}
\shortauthors{KILIC ET AL.}

\begin{document}

\title{Accurate Masses for the Primary and Secondary in the Eclipsing White Dwarf Binary NLTT 11748\footnote{Based
on observations obtained at the MMT Observatory, a joint facility of
the Smithsonian Institution and the University of Arizona.}}

\author{Mukremin Kilic\altaffilmark{1,4},
Carlos Allende Prieto\altaffilmark{2},
Warren R. Brown\altaffilmark{1},
M. A. Ag\"{u}eros\altaffilmark{3},
S. J. Kenyon\altaffilmark{1},
and Fernando Camilo\altaffilmark{3}}

\altaffiltext{1}{Smithsonian Astrophysical Observatory, 60 Garden St., Cambridge, MA 02138, USA; mkilic@cfa.harvard.edu}
\altaffiltext{2}{Instituto de Astrof\'{\i}sica de Canarias, 38205 La Laguna, Tenerife, Spain}
\altaffiltext{3}{Columbia University, Department of Astronomy, 550 West 120th Street, New York, NY 10027}
\altaffiltext{4}{\it Spitzer Fellow}

\begin{abstract}

We measure the radial velocity curve of the eclipsing detached white dwarf binary NLTT 11748. The primary
exhibits velocity variations with a semi-amplitude of 273 km s$^{-1}$ and an orbital period of 5.641 hr. We do not
detect any spectral features from the secondary star, or any spectral changes during the
secondary eclipse. We use our composite spectrum to constrain the temperature
and surface gravity of the primary to be $T_{\rm eff}= 8690 \pm 140$ K and $\log$ g = 6.54 $\pm 0.05$, which correspond
to a mass of 0.18 $M_{\odot}$. For an inclination angle of 89.9$^{\circ}$ derived from the eclipse modeling, the mass function
requires a 0.76 $M_{\odot}$ companion. The merger time for the system is 7.2 Gyr. However, due to the extreme mass ratio of
0.24, the binary will most likely create an AM CVn system instead of a merger.

\end{abstract}


\section{INTRODUCTION}

Radial velocity observations of extremely low-mass white dwarfs (0.2 $M_\odot$, ELM WDs) show that the majority are in
close binaries. This is expected, as the Galaxy is not old enough to produce such WDs through single star evolution.
Recent discoveries of seven short period binary WDs that contain ELM WDs increased interest in these systems \citep{kilic07,kilic09,kilic10,
vennes09,mullally09,marsh10,kulkarni10}.
Five of these systems will merge within a Hubble time, with the merger time being shorter than 500 Myr for three of them. The extreme
mass ratios of the binary components mean that some of these systems may not merge. Instead, they
may be the long-sought progenitors of AM CVn stars. On the other hand, depending on the inclination angle and the true mass ratio,
they may merge and create extreme helium stars, including R Coronea Borealis stars or single helium-enriched subdwarf O stars. If
the mass transfer is dynamically unstable, an underluminous Type Ia supernova is also a possibility \citep{guillochon10}.

\citet{kawka09} report the discovery of a nearby ELM WD in the New Luyten catalogue of stars with proper motions Larger than
Two Tenths of an arcsecond (NLTT). Based on low-resolution spectroscopy, they find that NLTT 11748 has $T_{\rm eff}=8540$ K,
$\log g=$ 6.2, and $M=0.167~M_\odot$. They estimate a distance of 199 pc. With a proper motion of 296.4 mas yr$^{-1}$ \citep{lepine05},
NLTT 11748 has a tangential velocity of 280 km s$^{-1}$. If NLTT 11748
were a single star, its kinematic properties would be similar to the runaway WD LP400$-$22 \citep{kilic09,vennes09,kawka06}.

To search for a binary companion, we obtained optical spectroscopy observations of NLTT 11748 in 2009 September. Subsequently,
\citet{steinfadt10}
reported the discovery of 3-6\% eclipses in the $g-$band light curve of this star and \citet{kawka10} 
presented an ephemeris and secondary mass function.
We use our spectroscopy data to confirm the period, search for spectroscopic signatures of the
secondary star during an eclipse, and also constrain the mass of the primary and secondary stars accurately.
Our observations are discussed in Section 2; the nature of the primary and the secondary are discussed in Section 3 and 4.

\section{OBSERVATIONS}

\subsection{MMT Optical Spectroscopy}

\citet{kawka09} reported the discovery of NLTT 11748 on 2009 September 17.
We obtained 52 spectra of NLTT 11748 with the 6.5 MMT and the Blue Channel
Spectrograph on UT 2009 September 26-28.
We used a 1$\arcsec$ slit and the 832 line mm$^{-1}$ grating in second order to obtain spectra with a wavelength
coverage of 3550 $-$ 4500 \AA, a resolving power of $R=$ 4300, and an exposure time of 450 s.
We obtained all spectra at the parallactic angle and acquired comparison lamp exposures either before or
after every science exposure.
We checked the stability of the spectrograph by measuring the centroid of the Hg emission line at 4358.34 \AA\
from street lights.
Over three nights, we measured an average offset of $-1.2 \pm 0.3$ km s$^{-1}$.
We flux-calibrated the spectra using the spectrophotometric standard BD+28 4211 \citep{massey88}.

To measure heliocentric radial velocities, we use the cross-correlation package RVSAO \citep{kurtz98}.
We obtain preliminary velocities by cross-correlating the observations with bright WD templates of known velocity.
However, greater velocity precision comes from cross-correlating the object with itself.
Thus we shift the individual spectra to rest-frame and sum them together into a high signal-to-noise ratio template spectrum.
Individual spectra have a signal-to-noise ratio of 30 in the continuum at 4000 \AA; the composite spectrum has a signal-to-noise
ratio of 200.
Our final velocities come from cross-correlating the individual observations with this template,
and are presented in Table 1.

We also use the best-fit WD model spectrum (see Section 3) to
measure radial velocities. The results are consistent within 5 km s$^{-1}$.
The mean velocity difference between the analyses is $3.0 \pm 0.6$ km s$^{-1}$.
Thus, the systematic errors in our measurements are $\approx$3 km s$^{-1}$.
This small uncertainty gives us confidence
that the velocities in Table 1 are reliable.

\subsection{Green Bank Telescope}

Without prior knowledge of the photometric eclipses discovered in this 
system, we targeted it for a millisecond pulsar companion search using the 
Green Bank Telescope \citep[see][]{agueros09a}. We observed NLTT 11748 
on 2010 Feb 07 for $1.4$ hr at $350$ MHz, which is where a pulsar with a 
typical spectral index of $\alpha = -1.6$ would be brightest, with the 
Green Bank Ultimate Pulsar Processing Instrument (GUPPI) 
backend.\footnote{https://wikio.nrao.edu/bin/view/CICADA/GUPPiUsersGuide} 
The data reduction was similar to that described in \citet{agueros09b}. 
We used the standard search techniques implemented in the PRESTO software 
package \citep{ransom01}. Because PRESTO assumes a constant apparent 
acceleration, we divided our GBT data into eight separate $644.25$~s 
integrations, each representing $\sim3\%$ of an orbit, and conducted 
searches for pulsations separately in each of these partial observations. 
Not surprisingly (given the 3-6\% eclipses detected in the optical light 
curve), no convincing pulsar signal is detected in our data.

\section{RESULTS}

The radial velocity of NLTT 11748 varies by as much as 564 km s$^{-1}$ between different observations,
revealing the presence of a companion object. 
We compare the radial velocities of the H$\gamma$ and H$\delta$ lines with the H8 and higher order Balmer lines; there are no
significant velocity differences between these lines. Hence, the observed Balmer lines are from only one star.
We weight each velocity by its associated error and solve for the best-fit
orbit using the code of \citet{kenyon86}. We estimate the errors in the orbital parameters
through Monte Carlo simulations of 10000 sets of radial velocities.
The heliocentric radial velocities are best fit
with a circular orbit and a radial velocity amplitude K = 273.4 $\pm$ 0.5 km s$^{-1}$.
The best-fit orbital period is 0.23503 $\pm$ 0.00013 d (5.641 hr) with spectroscopic conjunction
at HJD 2455100.855518 $\pm$ 0.000069.
Based on the photometric light curve and eight separate spectra, \citet{steinfadt10} measure an orbital period of 0.23506 d
and a velocity semi-amplitude of 271 $\pm$ 3 km s$^{-1}$. In addition, \citet{kawka10} measure an orbital period of 0.23506 d and a velocity
semi-amplitude of 274.8 $\pm$ 1.5 km s$^{-1}$. Our
measurements are consistent with these estimates.
Figure 1 shows the observed radial velocities and our best fit period for NLTT 11748.

NLTT 11748 is near the Taurus-Auriga molecular cloud and the 4430 \AA\ diffuse interstellar band is detected in
our high signal-to-noise ratio MMT spectrum (Figure 2). The observed 2-3\% absorption in this band corresponds
to an extinction of $E(B-V)\approx0.1$ \citep{krelowski87}, consistent with the \citet{kawka09} estimate.

We perform model fits to each individual spectrum and also to the composite spectrum using synthetic WD spectra kindly
provided by D. Koester. We use the 52 individual spectra to obtain a robust estimate of the errors in our analysis.
Figure 2 shows our model fits to the Balmer line profiles (top panel) and to the composite spectrum (middle panel).
We obtain a best-fit solution of
$T_{\rm eff}=8690 \pm 140$ K and $\log$ g = 6.54 $\pm 0.05$ from the observed composite spectrum. The best-fit
model does not match the observed spectrum in the blue, especially the higher order Balmer lines (middle panel).
Even in the fits to the observed line profiles (top panel), the lines remain poorly fit.
The bottom panel in Figure 2 shows the de-reddened spectrum against this best-fit model.
The line core strengths are overestimated in the models.
This problem is not evident in the \citet{kawka09} analysis, which uses lower resolution and lower
signal-to-noise ratio spectra limited to wavelengths longer than about 3800 \AA.

Contribution from a cool companion could dilute the line profiles. We search for the spectral signature of such a companion
using our spectroscopy during the secondary eclipse.
Based on our orbital fit, one of the 52 spectra was obtained at phase 0 (see Figure 1) and it covers an entire
185 s secondary eclipse. Based on the orbital
fit by \citet{steinfadt10}, another one of our 52 spectra covers the secondary eclipse. 
We derive $T_{\rm eff}=$ 8620 and 8490 K and $\log g=$ 6.53 and 6.43 from these two spectra, respectively.
These temperature and surface gravity estimates are consistent with $T_{\rm eff}= 8690 \pm 140$ K and $\log g = 6.54 \pm 0.05$
obtained from the composite spectrum.
Figure 3 shows these two spectra compared to the
average of the spectra
taken immediately before and after the secondary eclipse. We do not detect significant differences between these spectra in either case, indicating
that the secondary star does not significantly contribute to our spectrum. 
Hence, the observed line profiles cannot be explained by
contribution from a binary companion.

Calibrating fluxes to better than a few percent over wavelength ranges
spanning several hundred angstroms is a challenging task. Our flux calibration relies on the observations
of the standard star BD+28 4211. Even though observations of the other targets from the
same observing run do not show any flux calibration problems, NLTT 11748 is our brightest target and subtle systematic errors
may be important. Comparing
$T_{\rm eff}$ = 8690 $\pm$ 140 K and $\log g$ = 6.54 $\pm$ 0.05 from this study to the
$T_{\rm eff}$ = 8540 $\pm$ 50 K and $\log g$ = 6.20 $\pm$ 0.15 from \citet{kawka09} shows excellent
agreement between temperatures but a
systematic offset in gravity. Differences in the fitting
method employed (Kawka \& Vennes used H$\alpha$ to H9 while the present study
used H$\gamma$ to H14), model atmospheres, and flux calibration may all contribute toward a
systematic offset between gravity measurements which are very
sensitive to the strengths of the higher order Balmer lines. Fortunately, the differences in
surface gravity do not significantly impact the mass
derived from the mass-radius relations. 

Figure 4 shows the effective temperature and surface gravity for NLTT 11748 (filled circle) along with the previously
identified ELM WDs.
Comparing our temperature and surface gravity measurements to \citet{panei07} models \citep[updated by][]{kilic10}, NLTT 11748 has
$M\approx0.18~M_\odot$. Using their best-fit model spectra and the mass-radius relation of \citet{serenelli02},
\citet{kawka09} derive $M\approx0.17~M_\odot, M_V=9.7 \pm 0.3$ mag, $d=199 \pm 40$ pc (based on the 2MASS $J-$band photometry), and a
cooling age of 4-6 Gyr.
Based on the updated \citet{panei07} models, NLTT 11748 has $M_V = 10.28$ mag, $R = 0.038~R_\odot$,
and $d = 152 \pm 30$ pc. 

The orbital period and the semi-amplitude of the radial velocity variations imply a mass function of
0.4978 $\pm$ 0.0027. For $M=0.18~M_\odot$ and the inclination angle of
89.9$^{\circ}$ \citep{steinfadt10},
the companion is a 0.76 $M_\odot$ object at an orbital separation of 1.6 $R_\odot$.

\section{DISCUSSION}

Our radial velocity measurements show that NLTT 11748 is in a binary system with an orbital period of 5.641 hr.
Our best-fit model cannot perfectly match the high order Balmer line core strengths, however flux calibration is a
possible culprit.
Optical spectroscopy does not reveal any spectral features from a companion, and
the observed 3-6\% eclipses in the light curve \citep{steinfadt10} rule out main-sequence and
neutron star companions. A relatively cold ($T_{\rm eff}\leq 7400$ K) 0.76 $M_\odot$ C/O WD is the only solution
that satisfies the mass and radius constraints for the secondary star.

At a Galactic latitude of $-28.4^{\circ}$, NLTT 11748 is 48 pc below the plane. The observed systemic velocity
is 125.9 $\pm$ 0.4 (stat) $\pm$ 3.0 (sys) km s$^{-1}$;
the proper motion is
\citep[$\mu_{\alpha} cos \delta, \mu_{\delta}) = (236.1, -179.2$ mas yr$^{-1}$;][]{lepine05}.
Our systemic velocity measurement is lower than that of \citet{steinfadt10} and \citet{kawka10} and
the systematic errors dominate our velocity zero point. 
After correcting the systemic velocity for the gravitational redshift of 3 km s$^{-1}$,
the velocity components with respect to the local standard of rest as defined by \citet{hogg05} are
$U=-142 \pm 8, V=-187 \pm 41$, and $W=-29 \pm 6$ km s$^{-1}$. Clearly, NLTT 11748 is a halo star \citep[see also][]{steinfadt10,kawka10}.

NLTT 11748 is the only known eclipsing detached double WD system.
Modeling the optical light curve of the system,
\citet{steinfadt10} derive a radius of $\approx$ 0.038-0.040 $R_\odot$
for a 0.18 $M_\odot$ primary WD. These estimates are entirely consistent with the \citet{panei07} model predictions of 0.038 $R_\odot$
for a 8690 K, 0.18 $M_\odot$ WD. This result provides the first test of the theoretical mass-radius relations
for ELM WDs. The primary eclipse depth of 6.7\% implies that the radius of the C/O WD is 26\% of that of the ELM
WD, i.e. 0.0099-0.0104 $R_{\odot}$. This range is entirely consistent with the theoretically predicted
radii for 0.76 $M_{\odot}$ cool WDs \citep[$\approx$0.0105 $R_{\odot}$,][]{salaris10}.

The merger time due to loss of angular momentum through gravitational radiation is 7.2 Gyr.
If the mass transfer is dynamically unstable, the system merges to produce an extreme helium star or an underluminous
Type Ia supernova \citep{guillochon10}. However, with a mass ratio of 0.24, the mass transfer is probably stable
\citep[see][and references therein]{marsh04,nelemans10}.
NLTT 11748 is then one of the best known AM CVn progenitor candidates \citep[see also][]{kilic10}.
 
\section{CONCLUSIONS}

Using high signal-to-noise ratio medium-resolution spectroscopy, we improve the mass estimates for the primary
and secondary star in the eclipsing WD binary system NLTT 11748. We identify the visible component of the binary as a 8690 K,
0.18 $M_\odot$ WD at a distance of 152 $\pm$ 30 pc. 
The secondary is not detected in our MMT spectra. The mass
function for the system requires a 0.76 $M_\odot$ C/O core WD companion. 
Taking all three available mass functions (0.480, 0.505, and 0.498) from \citet{steinfadt10},
\citet{kawka10}, and this study and two available spectroscopic
mass estimates (0.17 and 0.18 $M_{\odot}$), we are confident that systematic errors
do not influence our interpretation.
The 3.5\% deep secondary eclipses constrain
the secondary to be relatively cool \citep[$T_{\rm eff}\leq7400$ K,][]{steinfadt10}. Follow-up time-series photometry
to detect the secondary eclipses in several different filters will be useful to constrain the temperature and WD cooling age
of the secondary star.

NLTT 11748 joins the growing list of short-period binary WDs including ELM WDs. Along with SDSS J0822+2753, J0849+0445,
and J1257+5428 \citep{kilic10,marsh10,kulkarni10}, NLTT 11748 is likely to form an AM CVn system due to its extreme mass ratio.

\acknowledgements
We thank the referee S. Vennes for helpful suggestions, D. Koester for kindly providing WD
model spectra, and S. Ransom for his help with the GBT observations.
Support for this work was provided by NASA through the {\em Spitzer Space Telescope} Fellowship Program,
under an award from Caltech. M.A.A.\ is supported by an NSF Astronomy and Astrophysics Postdoctoral 
Fellowship under award AST-0602099. The Robert C.\ Byrd Green Bank Telescope is operated
by the National Radio Astronomy Observatory, which is a facility of the US National Science Foundation
operated under cooperative agreement by Associated Universities, Inc.

{\it Facilities:} \facility{MMT (Blue Channel Spectrograph)}, \facility{Green Bank Telescope}

\begin{deluxetable}{cr}
\tablecolumns{2}
\tablewidth{0pt}
\tablecaption{Radial Velocity Measurements for NLTT 11748}
\tablehead{
\colhead{HJD}&
\colhead{Heliocentric}\\
+2455100 & Radial Velocity (km s$^{-1}$)
}
\startdata
0.937114 & 350.0 $\pm$ 2.8 \\
0.942508 & 329.5 $\pm$ 3.2 \\
0.950136 & 280.4 $\pm$ 2.8 \\
0.955842 & 246.0 $\pm$ 2.6 \\
0.961885 & 205.9 $\pm$ 2.1 \\
0.969501 & 148.0 $\pm$ 3.7 \\
0.974895 & 110.5 $\pm$ 2.4 \\
0.980289 & 71.7 $\pm$ 3.5 \\
0.987975 & 13.9 $\pm$ 3.7 \\
0.993369 & $-$15.0 $\pm$ 2.8 \\
0.998763 & $-$46.6 $\pm$ 3.8 \\
1.005962 & $-$79.5 $\pm$ 3.4 \\
1.011345 & $-$90.3 $\pm$ 3.9 \\
1.016739 & $-$116.8 $\pm$ 4.7 \\
1.021345 & $-$150.9 $\pm$ 7.5 \\
1.880303 & 336.5 $\pm$ 3.4 \\
1.932877 & $-$6.2 $\pm$ 2.5 \\
1.938271 & $-$47.5 $\pm$ 4.2 \\
1.943746 & $-$79.0 $\pm$ 2.7 \\
1.951455 & $-$113.0 $\pm$ 2.7 \\
1.956849 & $-$124.9 $\pm$ 2.9 \\
1.962243 & $-$136.4 $\pm$ 3.1 \\
1.969477 & $-$153.3 $\pm$ 3.0 \\
1.980265 & $-$134.9 $\pm$ 3.8 \\
1.985694 & $-$127.7 $\pm$ 2.4 \\
1.992939 & $-$113.9 $\pm$ 3.0 \\
1.998333 & $-$92.4 $\pm$ 4.7 \\
2.003716 & $-$60.2 $\pm$ 2.9 \\
2.009110 & $-$18.4 $\pm$ 2.7 \\
2.014909 & 6.9 $\pm$ 3.4 \\
2.018740 & 44.1 $\pm$ 4.8 \\
2.022919 & 78.9 $\pm$ 6.9 \\
2.026403 & 99.5 $\pm$ 7.0 \\
2.928187 & $-$116.3 $\pm$ 3.6 \\
2.933581 & $-$101.4 $\pm$ 3.0 \\
2.938975 & $-$74.1 $\pm$ 2.4 \\
2.946615 & $-$40.6 $\pm$ 3.0 \\
2.952009 & $-$9.4 $\pm$ 2.7 \\
2.957391 & 40.3 $\pm$ 2.6 \\
2.962785 & 71.4 $\pm$ 4.5 \\
2.970274 & 111.1 $\pm$ 2.5 \\
2.975656 & 148.4 $\pm$ 4.0 \\
2.981050 & 203.9 $\pm$ 3.1 \\
2.986444 & 247.0 $\pm$ 2.9 \\
2.993898 & 276.2 $\pm$ 4.6 \\
2.999292 & 300.1 $\pm$ 2.9 \\
3.004686 & 348.6 $\pm$ 2.8 \\
3.010069 & 368.6 $\pm$ 2.4 \\
3.015416 & 381.3 $\pm$ 3.7 \\
3.018484 & 390.1 $\pm$ 4.8 \\
3.021551 & 382.1 $\pm$ 5.1 \\
3.024618 & 410.6 $\pm$ 5.7
\enddata
\end{deluxetable}

\begin{figure}
\includegraphics[width=5in,angle=-90]{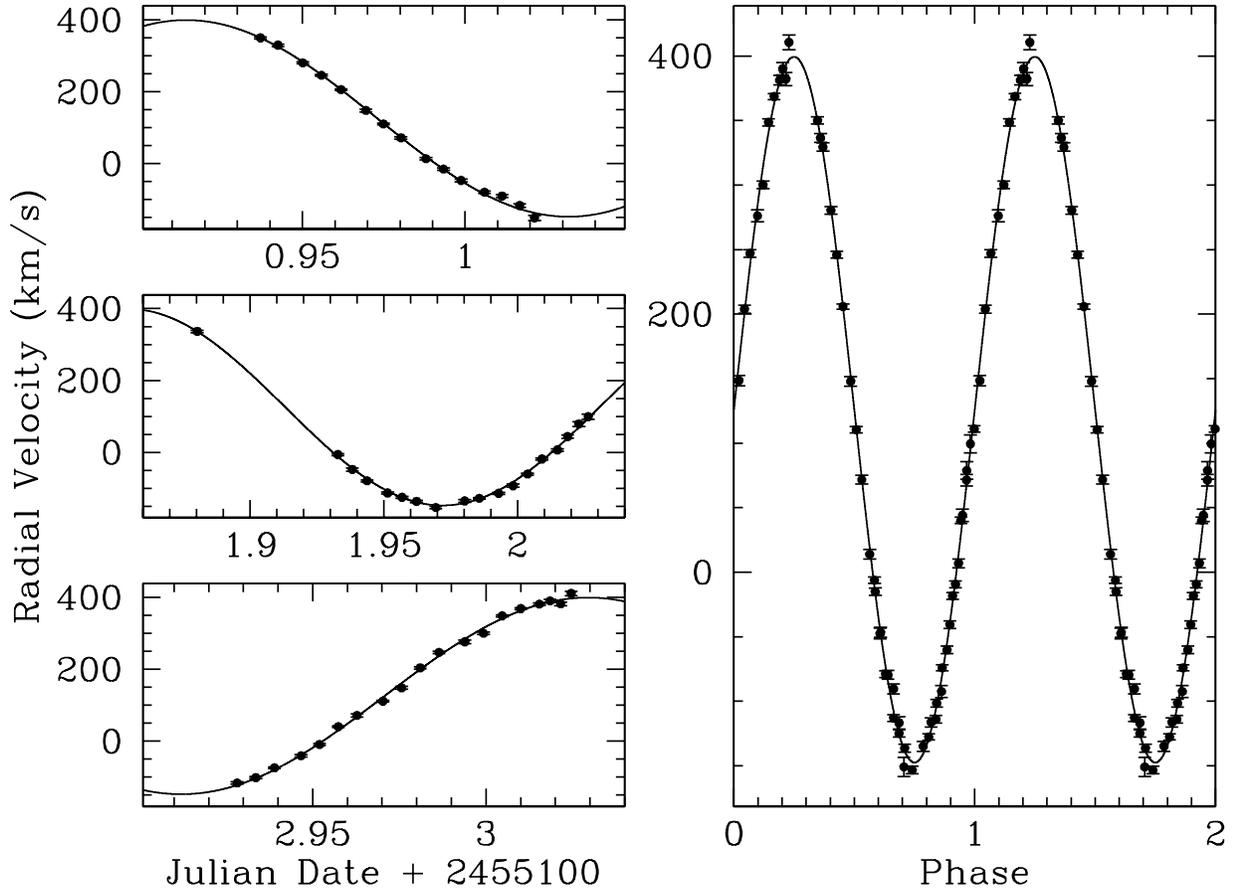}
\caption{Radial velocity of NLTT 11748 (black dots) observed in 2009 September (left panels).
The right panel shows all of these data points phased with the best-fit period. The solid line represents the best-fit
model for a circular orbit with a radial velocity amplitude of 273.4 km s$^{-1}$ and a period of 0.23503 days.}
\end{figure}

\begin{figure}
\includegraphics[width=6in,angle=0]{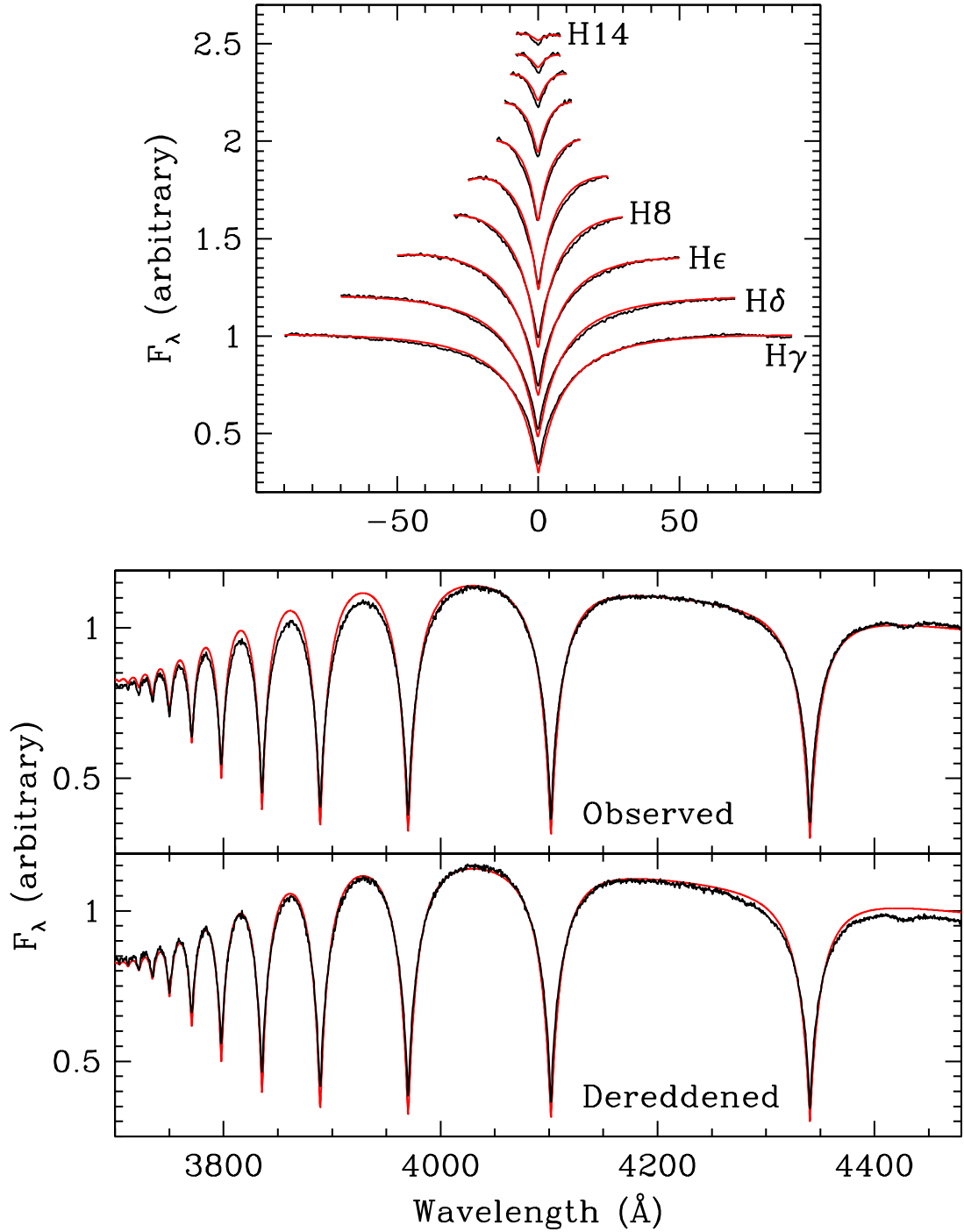}
\caption{Spectral fits (solid lines) to the flux-normalized line profiles (jagged lines, top panel) and
to the observed composite spectrum of NLTT 11748 (middle panel).
The bottom panel shows the de-reddened spectrum.}
\end{figure}

\begin{figure}
\includegraphics[width=5in,angle=-90]{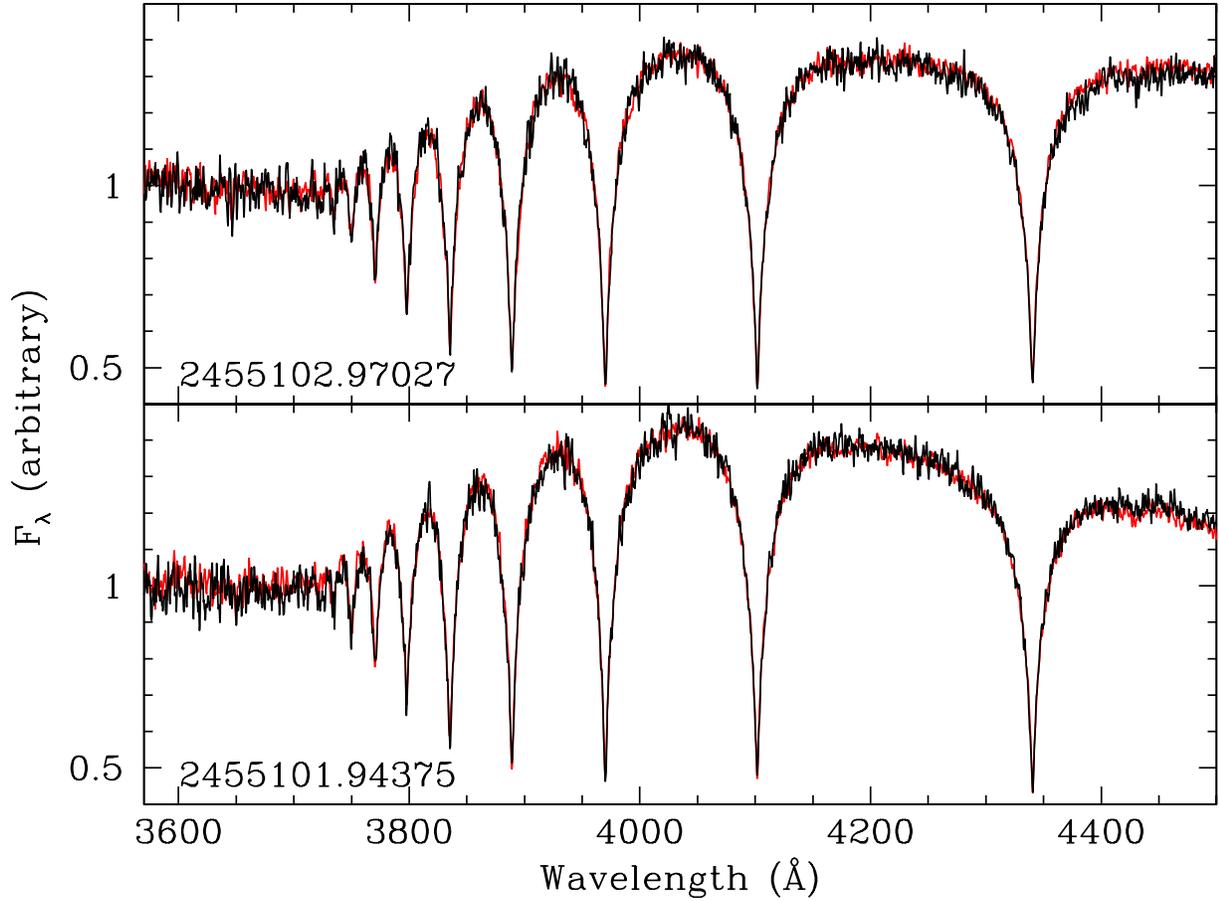}
\caption{Secondary eclipse spectrum of NLTT 11748 (black line) compared to the average spectra taken
immediately before and after the eclipse (red line). The top panel is for our ephemeris determination and the bottom
panel is for the ephemeris found by \citet{steinfadt10}. The mid-exposure HJD is given in each panel.
No evidence of the companion is seen during secondary eclipse.}
\end{figure}

\begin{figure}
\plotone{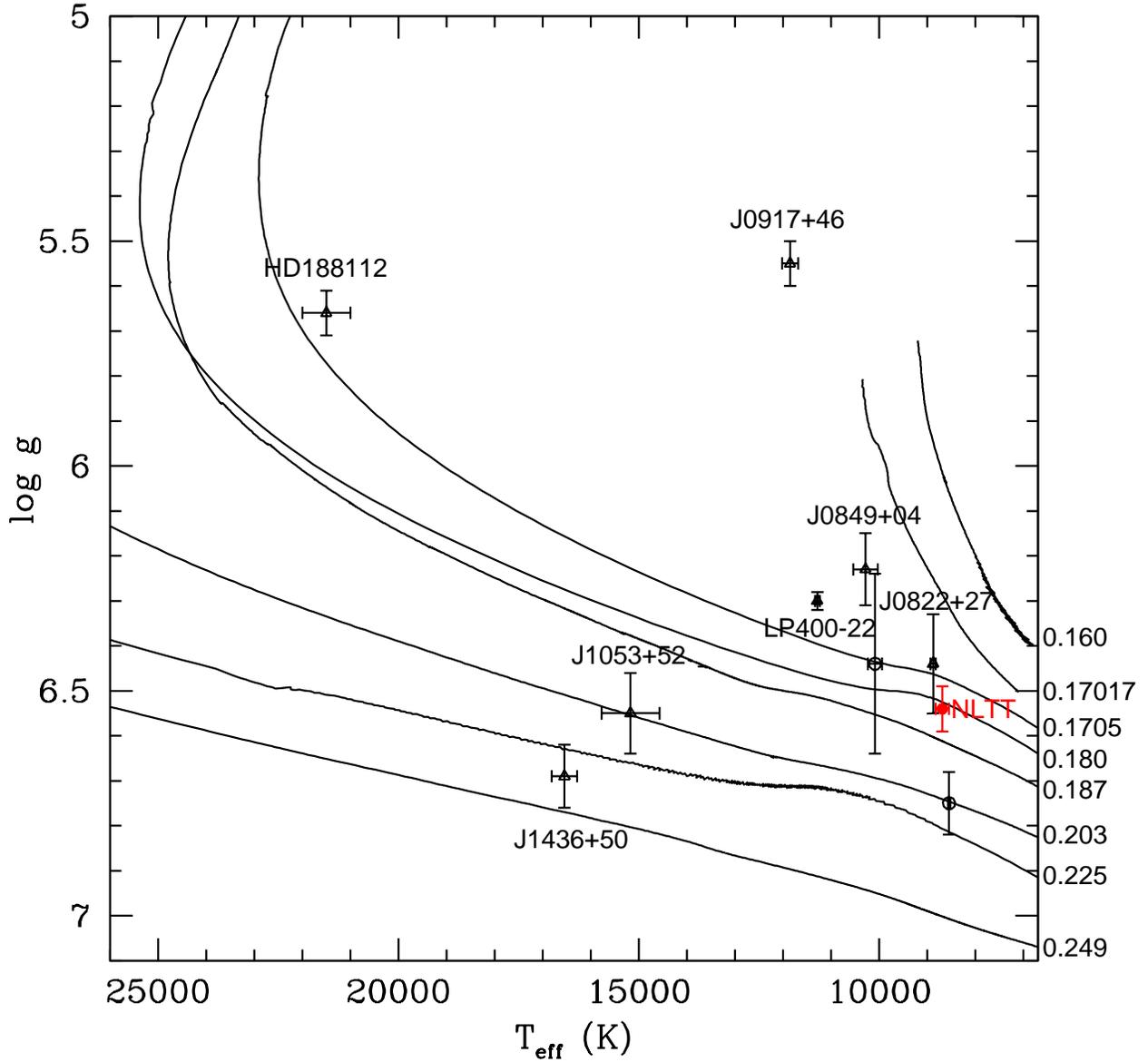}
\caption{The best fit solution for the surface gravity and temperature of NLTT 11748 (filled circle),
overlaid on tracks of constant mass from \citet[][based on the Panei et al. 2007 models]{kilic10}.
Spectroscopically confirmed ELM WDs in the literature \citep[see][]{kilic10} and the subdwarf B star HD 188112 \citep{heber03}
are shown as open symbols.}
\end{figure}

\end{document}